\documentclass{aastex631}

\begin{document}

\title{Understanding the Travel-time Asymmetry of Acoustic Waves in Sunspots With Time–distance Helioseismology}%

\author{Haiyu Li}
\affiliation{School of Earth and Space Sciences, Peking University, Beijing, 100871, People's Republic of China; haiyuli@stu.pku.edu.cn; huitian@pku.edu.cn}
\affiliation{State Key Laboratory of Solar Activity and Space Weather, National Space Science Center, Chinese Academy of Sciences, Beijing, 100190, China}
%%\email{haiyuli@stu.pku.edu.cn}

\author{Tobías Felipe}
\affiliation{Instituto de Astrofísica de Canarias, c/vía Láctea, s/n, E-38205 La Laguna, Tenerife, Spain}
\affiliation{Departamento de Astrof\'isica, Universidad de La Laguna, E-38205 La Laguna, Tenerife, Spain}

\author{Elena Khomenko}
\affiliation{Instituto de Astrofísica de Canarias, c/vía Láctea, s/n, E-38205 La Laguna, Tenerife, Spain}
\affiliation{Departamento de Astrof\'isica, Universidad de La Laguna, E-38205 La Laguna, Tenerife, Spain}

\author{Hui Tian}
\affiliation{School of Earth and Space Sciences, Peking University, Beijing, 100871, People's Republic of China; haiyuli@stu.pku.edu.cn; huitian@pku.edu.cn}
\affiliation{State Key Laboratory of Solar Activity and Space Weather, National Space Science Center, Chinese Academy of Sciences, Beijing, 100190, China}
%%\email{huitian@pku.edu.cn}

\author{Paul Rajaguru}
\affiliation{Indian Institute of Astrophysics, Bangalore 560034, India}

\author{Yuhang Gao}
\affiliation{School of Earth and Space Sciences, Peking University, Beijing, 100871, People's Republic of China; haiyuli@stu.pku.edu.cn; huitian@pku.edu.cn}
\affiliation{Centre for mathematical Plasma Astrophysics, Department of Mathematics, KU Leuven, Celestijnenlaan 200B bus 2400, B-3001 Leuven, Belgium}

\begin{abstract} Mapping the subsurface structure and flow field of sunspots has been a challenging task for helioseismology. In this work, we investigate the propagation of acoustic waves in a sunspot in NOAA active region 11312 using time--distance helioseismology. Travel times of waves traveling into and out of the sunspot are measured as functions of travel distance and azimuthal angle relative to the local radial direction. The same time--distance analysis is also applied to a simulated data based on a magnetohydrostatic (MHS) model of sunspot, and forward modeling of travel times is performed using ray tracing based on both the MHS sunspot model and a magnetohydrodynamic (MHD) simulation. We find that both ingoing (traveling from the quiet area into the sunspot) and outgoing waves (traveling from the sunspot into the quiet area) have shorter travel times than in the quiet Sun, with travel-time reductions of up to $\sim$40~s. The magnitude of the mean time shift is largest for waves traveling along the radial direction at small travel distances. A clear asymmetry is detected between ingoing and outgoing waves: outgoing waves generally exhibit shorter travel times. This asymmetry is strongest for radial direction and small travel distances, with differences exceeding 1 min for 3.5~mHz and 4.5~mHz waves. From the results of both observations and models, our analysis indicates that the overall reduction in travel time could be primarily caused by the combined effects of Wilson depression, magnetic field, and wave-speed perturbations, while the ingoing--outgoing asymmetry could be partly attributable to subsurface flows. Although the forward-modeling results reproduce several qualitative features of the observations, quantitative discrepancies remain, highlighting limitations of current sunspot models and ray-theoretical approximations. \end{abstract}


\section{Introduction} \label{sec:intro}

Time--distance helioseismology \citep{Duvall:1993aa,Duvall:1996aa} has long been used to investigate the subsurface structure and dynamics of sunspots. Early applications revealed the presence of large-scale flows beneath sunspots, including near-surface converging flows and outflows in deeper layers \citep{Kosovichev:1996aa,Zhao_2001}. %
Subsequent studies using f-mode and p-mode travel-time measurements further characterized the moat flow and its depth dependence \citep{Duvall:2000aa,Gizon:2009aa}. Inversions of time--distance measurements have also enabled estimates of subsurface kinetic helicity, showing consistency with magnetic helicity inferred from vector magnetograms \citep{Zhao:2003aa,Gao_2012}. 
In parallel, other local helioseismology techniques such as helioseismic holography, acoustic imaging, and ring-diagran analysis, also revealed similar flow structures \citep[e.g.,][]{LindseyC.1996,Sun:1997aa,Haber:2000aa,Haber:2004aa,Hindman_2009}. 
Collectively, these results support the existence of organized flow structures beneath sunspots, typically confined to the upper $\sim$10~Mm of the convection zone, although the inferred flow magnitudes vary substantially between studies \citep{Kosovichev2010LocalHO}.

The subsurface thermal and magnetic structures of sunspots have also been extensively investigated. Employing ray-approximation sensitivity kernels, \citet{Kosovichev:1996aa} inverted for the first tomographic maps of wave-speed\footnote{Since in sunspots the speed of waves is also affected by the magnetic field, we choose to call the wave propagating speed in reversion as \textit{wave speed}, and the thermal acoustic speed in theoretical analysis and modeling as \textit{sound speed} ($c_\mathrm{s}=\sqrt{\Gamma p/\rho}$ in \ref{subsec:The Eikonal Method}).} perturbations beneath sunspots. Later refinements revealed a two-layer structure consisting of a shallow region of reduced wave speed overlying a deeper region of enhanced wave speed \citep[e.g.,][]{Kosovichev:2000aa,Jensen:2001aa}. However, to what extent this general picture is accurate is still under debate due to other factors besides the wave-speed perturbation that may affect the travel-time measurements \citep{Gizon:2009aa}. Moreover, using time-distance helioseismology, \citet{10.1093/mnras/stv506} revealed that sunspot magnetic field strength could cause frequency-dependent directional behaviour of acoustic wave travel-time shifts, and other effects including the Wilson depression may also significantly affect the wave travel times. 

To study the sunspot's subsurface structure, multiple sunspot models for helioseismic study were constructed \citep{Khomenko_2008,Rempel:2009sci,Cameron:2011aa}, and numerical modeling based on these idealized or numerical sunspot models has provided complementary insight. For example, \citet{Birch_2004} used their modeling results to show that details such as in time-distance measurements filtering can affect the inversion results. Case studies of the sunspot in NOAA Active Region (AR) 9787 \citep{Liang2013,Schunker2013} combining model simulation with measurements suggested that thermal structures including the Wilson depression can cause detectable time shifts. The effect of the Wilson depression is also discussed by \citet{Felipe2017}. \citet{Zhao:2020aa} reported another systematic ``center-to-limb like" effect causing travel time difference of waves traveling between different heights of sunspot, resembling center-to-limb effect of the Sun. 
Recently, \citet{Duvall2018} performed two-skip time–distance helioseismology measurements, and emphasized the impact of thermal structures in their measurements. Regarding the role of magnetic fields in modifying wave propagation, numerous studies have demonstrated that near-surface magnetic fields can strongly affect acoustic waves through mode conversion near the layer where sound speed $c_{\mathrm{s}}$ equals the Alfv\'en speed $v_{\mathrm{A}}$ \citep[e.g.,][]{Couvidat:2007aa,Khomenko_2009,Moradi:2008aa}. Theoretical efforts \citep{10.1093/mnras/stt1473,Schunker&Cally2006MNRAS} also revealed the impact of magnetic field on mode conversion and phase shifts, depending on the field's strength and inclination. 

In summary, magnetic field is expected to increase wave speed and shorten travel times; Wilson depression changes the location of the cutoff layers and probably leads to shorter travel times; flows can act differently in ingoing and outgoing directions. Although the impacts of magnetic and thermal structures had been widely investigated, disentangling the relative contributions of magnetic fields, flows, and thermal perturbations to observed travel-time shifts remains a major challenge.

In this work, we apply time--distance helioseismology measurements on a well-isolated sunspot in AR 11312. We measure travel times of waves traveling into and out of the sunspot over a range of horizontal distances and azimuthal angles relative to the sunspot's local radial direction, which helps us to distinguish the contributions of different factors. To interpret the measurements, we analyze simulated Doppler data generated from a magnetohydrostatic (MHS) sunspot model \citep{Felipe_2016} and perform ray-tracing forward modeling using both this model and a magnetohydrodynamic (MHD) sunspot model \citep{Rempel_2009apj,Rempel:2009sci,Braun:2012aa}. The observational analysis is described in Section~\ref{sec:Observations and time-distance measurement}, the forward modeling calculations are presented in Section~\ref{sec:simulation}, and the results are summarized and discussed in Section~\ref{sec:Summary}.

\section{Observations and Time--Distance Measurements} \label{sec:Observations and time-distance measurement}

The sunspot analyzed in this study was located in AR 11312. During the observing period the sunspot exhibited an approximately circular morphology and remained relatively stable in size and structure. We also examined the vector magnetic field observations of the sunspot and found that the horizontal field is mostly along the sunspot radial direction. The observations spanned a total of five days, from 2011 October 8 00:00 UT to October 12 24:00 UT, during which the sunspot rotated from about $33\degr$ east side of the central meridian to about $33\degr$ west side of the central meridian. We use Doppler velocity observations of Fe \textsc{i} 6173$\text{\AA}$ line obtained by the Helioseismic and Magnetic Imager (HMI; \citealt{Scherrer:2012aa,Schou:2012aa}) onboard the \textit{Solar Dynamics Observatory} ($SDO$; \citealt{Pesnell:2012aa}). The entire observational period is divided into fifteen continuous 8-hour data cubes with a temporal cadence of 45~s. Each Dopplergram is remapped into Postel projection and tracked at the Carrington rotation rate. The resulting data cubes have a spatial size of $512\times512$ pixels with a plate scale of 0.03 heliographic degrees per pixel, corresponding to approximately 0.365~Mm per pixel. For comparison, we also analyze six 8-hr datasets of a quiet-Sun region observed between 2011 October~2 and October~3. These quiet-Sun measurements are prepared in the same way as the sunspot data and serve as a reference for isolating travel-time perturbations associated with the sunspot. Prior to the helioseismic analysis, a running‑difference is applied to each Dopplergram to suppress low‑frequency (lower than 1.5 mHz) convective signals and surface gravity ($f$) mode, but retain acoustic ($p$‑mode) oscillations.

\subsection{Measurement Schemes} \label{subsec:Data Analysis Method}

\begin{figure}[ht!]
\centering
\resizebox{0.70\textwidth}{!}{\plotone{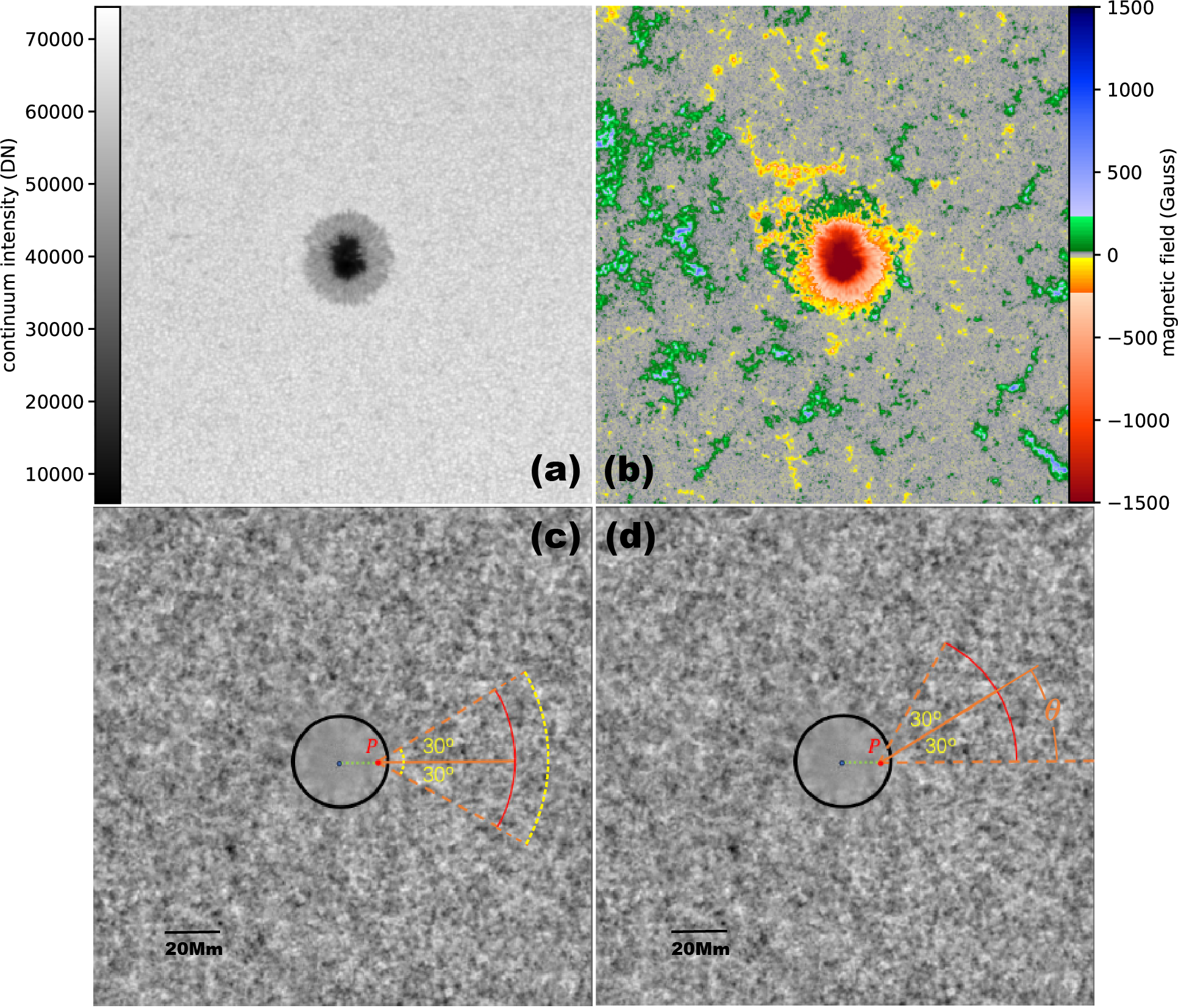}}
\caption{Upper panels: HMI continuum intensitygram (panel a) and magnetogram (panel b) of the sunspot, taken at 00:00 UT on 2011 October 11. Lower panels: filtered Dopplergram of the sunspot with schematic illustrations of the time--distance measurement geometry. The black circle indicates the boundary of selected sunspot area. For a given pixel $P$ inside the sunspot area, acoustic travel times are measured between $P$ and a circular arc located in the surrounding quiet area. 
The local radial direction is defined as the line connecting the sunspot center to $P$ and corresponds to $\theta=0^{\circ}$ (panel c).
The arc subtends an angular extent of $60^{\circ}$ and is rotated in steps of $30^{\circ}$ to sample different traveling directions. To better understand the measurement scheme of different directions, the $\theta=30^{\circ}$ case is shown in panel d. Arcs with maximum and minimum radius (correspond to maximum and minimum travel distances) in this work are shown as dashed yellow curves in panel c. Only arcs fully located outside the sunspot boundary are used to ensure correlations between sunspot and quiet-area signals.
\label{fig:sunspot1}}
\end{figure}

The goal of this study is to investigate how acoustic-wave travel times vary with traveling direction relative to the local radial direction of the sunspot. In particular, we aim to measure travel times of waves traveling into the sunspot (ingoing waves) and out of the sunspot (outgoing waves), and to determine how these travel times depend on distance, wave frequency and azimuthal angle.

In time–distance helioseismology, wave travel times are commonly measured from the temporal cross-covariance function of oscillation signals between two spatial locations \citep{Duvall:1993aa}. The cross-covariance function represents the average wave packet traveling between two spatial locations and allows the travel time of acoustic waves to be determined. However, because the signal-to-noise ratio of helioseismic measurements for a single pair of pixels is generally low, it is standard practice to average signals over extended spatial regions (e.g., arcs or annuli) in order to enhance the coherent wave signal. In this work, we adopt an arc-averaging scheme that allows directional measurements of travel times relative to the sunspot radial direction. Positive time lags of a cross-covariance function correspond to waves traveling from a point inside the sunspot to an averaged arc in surrounding quiet Sun (outgoing waves), while negative time lags correspond to waves traveling from the arc in the quiet Sun toward the point in the sunspot (ingoing waves). Definition of shape of the arc will be described in the following paragraphs.

For each 8-hr data segment, we define a circular region encompassing the sunspot, including its umbra and penumbra. Because the sunspot maintains a nearly constant size and shape over the observing period, a fixed circle with a radius of 48 pixels ($\approx17.5$~Mm) is used for all datasets. We refer to this region as the \emph{sunspot area}, while the surrounding region outside this boundary is referred as the \emph{quiet area} (This quiet area should be distinguished from the independent quiet-Sun region that is used for control measurements). The network magnetic fields in the quiet area are weak and we neglect their contribution to travel times.

A schematic illustration of the measurement geometry is shown in Figure \ref{fig:sunspot1}. For each pixel $P$ within the sunspot area, we measure travel times between that pixel and surrounding quiet-area locations (ingoing and outgoing). The local radial direction at $P$ is defined as the line connecting the sunspot center to that pixel. This direction is taken as the reference direction (azimuthal angle $\theta = 0^{\circ}$) for the measurement of directional travel times.

To measure wave traveling along direction $\theta = 0^{\circ}$, we construct a circular arc centered on pixel $P$. The arc subtends an angular extent of $60^{\circ}$ and its apex initially aligns with the local radial direction, shown as a solid red arc in Figure~\ref{fig:sunspot1}c. The arc radius ranges from 20 to 150 pixels (7.3--54.7~Mm, yellow dashed arcs in Figure~\ref{fig:sunspot1}c), corresponding to different horizontal travel distances $r$. For each arc radius, the arc is required to lie entirely outside the sunspot area's boundary. This constraint ensures that the oscillation signals used in the cross-correlation are obtained from the quiet area, so that the resulting travel times correspond to waves traveling between the sunspot and the surrounding quiet Sun. If the arc intersects the boundary for a given radius, the radius is increased until the entire arc falls within the quiet area. 

In order to capture the travel time signals, cross-covariance computation is performed at each horizontal travel distance $r$. For each pixel $P$ and each arc radius, we compute the temporal cross-covariance function between the Doppler velocity time series at pixel $P$ and the spatially averaged Doppler signal over all pixels along the arc:

\begin{equation}
C(\mathbf{x},\mathbf{d},\tau)=
\frac{1}{T}
\int_{0}^{T}
\psi(\mathbf{x},t)
\bar{\psi}(\mathbf{x}+\mathbf{d},t+\tau)
\, dt ,
\end{equation}

\noindent where $\psi$ denotes the Doppler velocity signal after performing the running-difference filtering, $\mathbf{x}$ is the location of the central pixel $P$, $\mathbf{d}=(d,\theta)$ is the horizontal displacement between the central pixel and the arc apex in the direction of the azimuthal angle $\theta$, $\tau$ is the time lag, and $T$ is the total duration of the data cube. In particular, the signal $\bar{\psi}(\mathbf{x}+\mathbf{d},\tau)$ represents the spatially averaged Doppler signal over all pixels along the selected arc. This cross-covariance function represents the average wave packet propagating between the sunspot pixel and the surrounding arc. The procedure is repeated for all pixels within the sunspot area. The resulting cross-covariance functions are then averaged over all such pixels to improve the signal-to-noise ratio. 

To sample different traveling directions $\theta$, at each pixel, the arc is rotated counterclockwise in steps of $30^{\circ}$. This procedure yields twelve directions corresponding to azimuthal angles $\theta = 0^{\circ}, 30^{\circ}, \dots, 330^{\circ}$ relative to the local radial direction. The process described above is applied for all twelve cases, producing a time–distance diagram for each direction (12 diagrams in total). Each diagram shows the cross-covariance as a function of travel time and horizontal distance at given direction.

For the quiet-Sun datasets, which are assumed to be statistically isotropic, the temporal cross-covariance is computed using full circular annuli instead of an arc, yielding another time–distance diagram for reference.

\subsection{Travel-time Shift Measurements}\label{Travel-time Measurements}

\begin{figure}[ht!]
\plotone{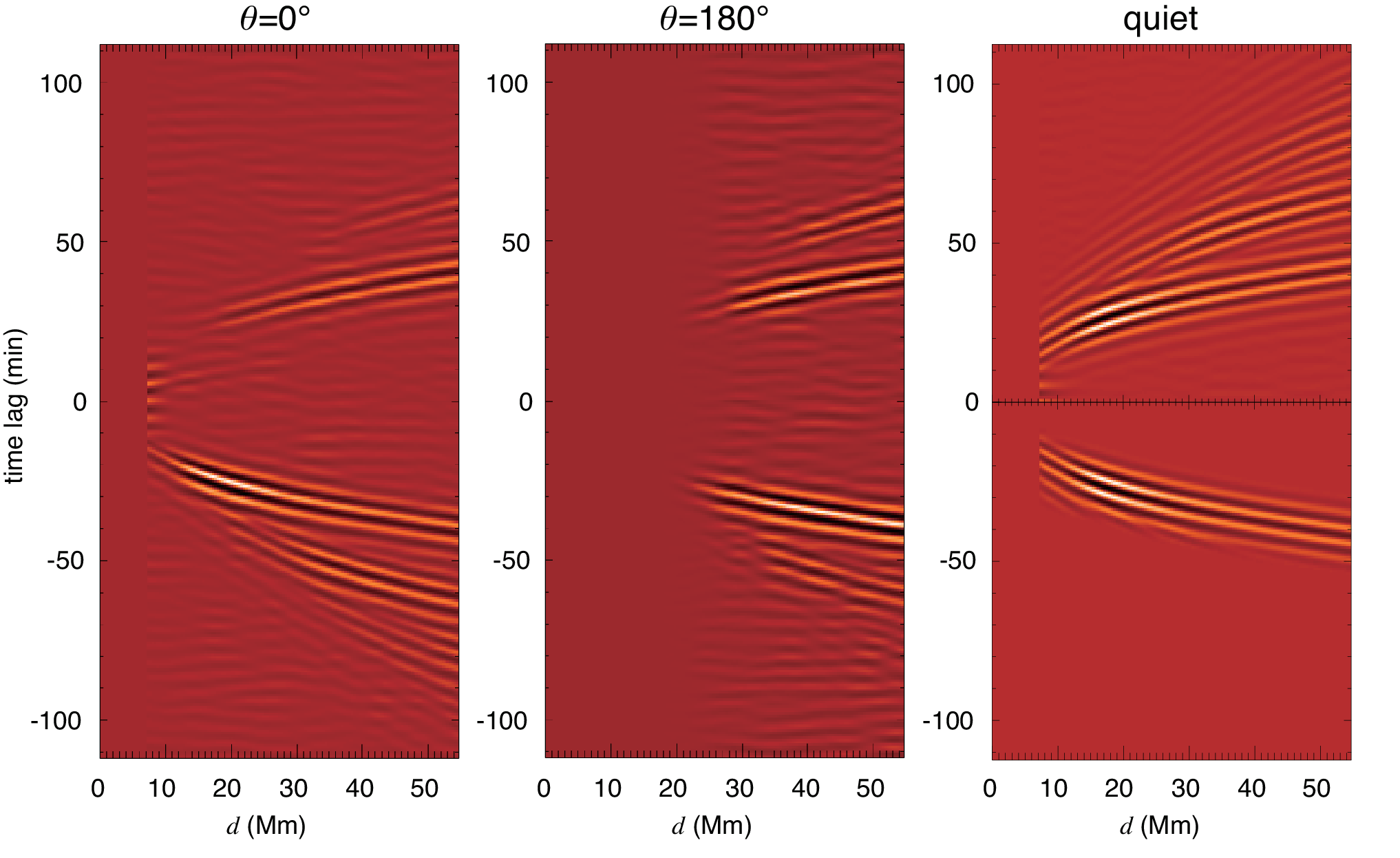}
\caption{Example time--distance diagrams obtained from whole time Doppler velocity series. Left panel: time--distance diagram at direction $\theta=0$°, middle panel: time--distance diagram at direction $\theta=180$°, right panel: time--distance diagram of quiet region with the positive time lag representing original cross-covariance and the negative time lag after applying a cosine-bell filter. The color scale represents the temporal cross-covariance as a function of time lag and horizontal distance, brighter color indicates stronger correlation. The first-skip acoustic branch used for travel-time measurements is clearly visible and is isolated using a temporal cosine-bell window.
\label{fig:tddiagram}}
\end{figure}

%The time-distance diagrams of $\theta=0$, 180° cases and the quiet region are shown in 
Figure \ref{fig:tddiagram} shows representative unfiltered time–distance diagrams for traveling directions of $\theta=0$° and $180$°, together with the corresponding quiet-Sun result. The quiet-Sun time–distance diagram is symmetric with respect the 0 time lag axis. In the diagrams, wave packets traveling between the central pixel and the surrounding arc appear as branches corresponding to different acoustic wave paths. Positive time lags correspond to waves traveling from the central pixel toward the surrounding arc (outgoing waves), and negative time lags correspond to waves traveling from the arc toward the central pixel (ingoing waves). The time–distance diagrams typically contain multiple branches corresponding to acoustic waves undergoing different numbers of surface reflections (or “bounces”). In this work we focus on the first-skip branch, which corresponds to waves that travel directly between the two locations without surface reflections, which samples a deeper layer without interacting with intermediate surface structures, providing robust travel-time measurements. To isolate this branch, we multiply a cosine-bell temporal window function centered on the middle of first-skip branch at each traveling distance. The window function at each distance has the amplitude of 1 for the first-skip branch and the width of 15~min, and the amplitude drops to 0 in the form of a cosine function at both ends of the branch, each dropping part has the width of 7.5~min. This filtering suppresses contributions from other branches and reduces contamination from noise. The filtered quiet-Sun diagram is shown in the region with negative time lag of right panel in Figure \ref{fig:tddiagram}.

Following the approaches by \citet{Chen_2018} and \citet{10.1093/mnras/staf309}, we calculate the travel-time shifts through measuring the phase difference between two cross-covariance functions in the frequency domain. Let $C_{\mathrm{spot}}(d,\tau)$ and $C_{\mathrm{QS}}(d,\tau)$ denote the filtered cross-covariance functions for the sunspot region and the quiet-Sun reference region, respectively. Using the temporal Fourier transforms $\widetilde{C}(\nu)=\mathcal{F}\{C(\tau)\}$ where $\nu$ is frequency, the phase shift between the two signals is obtained from the argument of their cross-spectrum,

\begin{equation}
\delta \phi(d,\nu)
=
\arg
\left[
\widetilde{C}_{\mathrm{spot}}(d,\nu)
\,
\widetilde{C}_{\mathrm{QS}}^{\dagger}(d,\nu)
\right],
\end{equation}

\noindent where the superscript $^{\dagger}$ denotes complex conjugation. The corresponding frequency-dependent travel-time shift is then given by

\begin{equation}
\delta \tau(d,\nu)
=
\frac{\delta \phi(d,\nu)}{2\pi\nu}.
\end{equation}

This quantity represents the difference between travel-times of waves traveling through the sunspot and those traveling through the quiet Sun. Because the cross-covariance function contains both positive and negative time-lag branches, phase shifts (and travel-time shifts) for ingoing and outgoing waves are measured separately by applying the above procedure to the corresponding parts of the cross-covariance function.

\subsection{Results of Measurements}\label{subsec:Result of Measurement}

\begin{figure}[htb!]
\centering
\resizebox{1.0\textwidth}{!}{\plotone{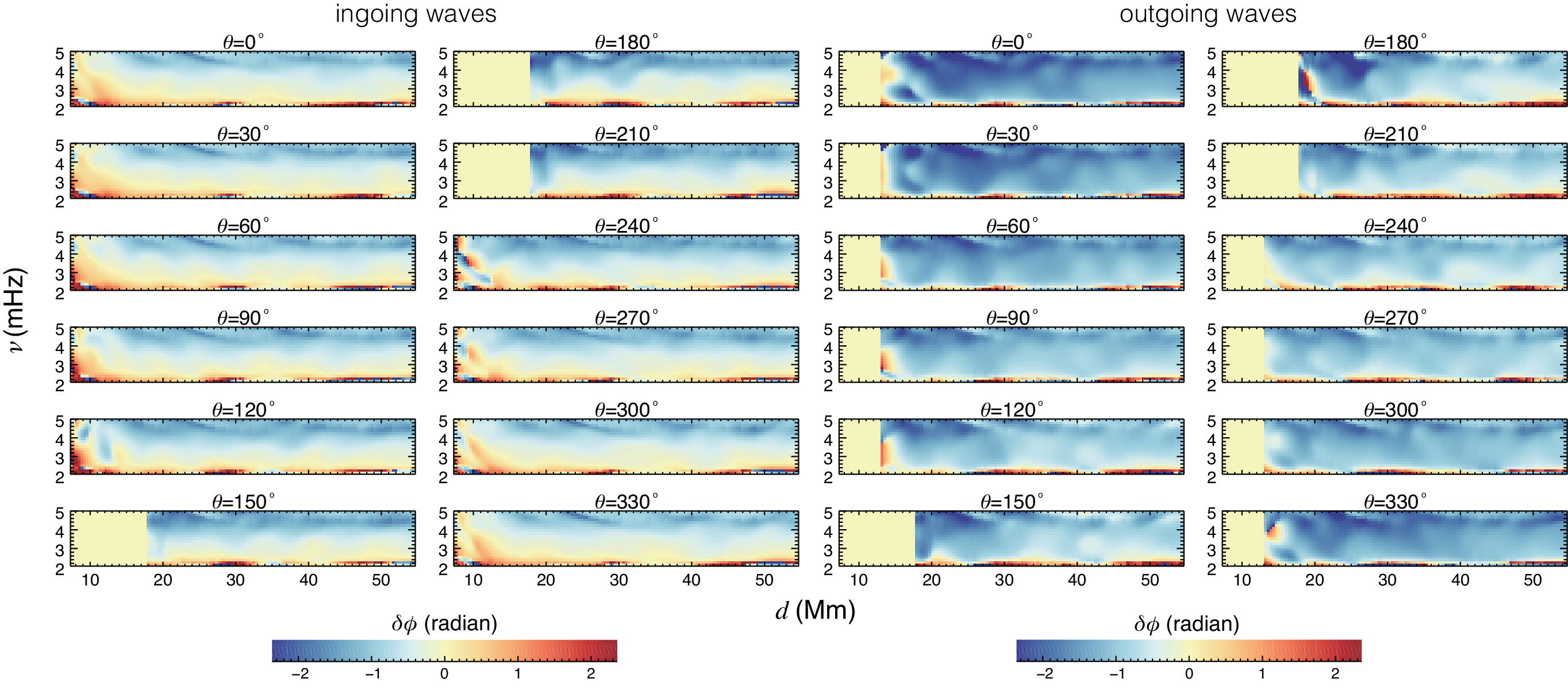}}
\caption{Measured phase shifts of ingoing (two left columns) and outgoing (two right columns) acoustic waves for different traveling angle $\theta$ and distance $d$. $\delta \phi$ is computed relative to the quiet-Sun region. Each panel displays results as a function of distance (horizontal axis) and frequency (vertical axis). The color scale ranges from $-0.75\pi$ to 0.75$\pi$ radians, with negative values indicating shorter travel times compared to the quiet Sun. %For directions $\theta$ = 150°, 180°, 210°, and other outgoing waves data at short distances are excluded due to poor signal quality. These omitted values are shown as zero. 
$\delta \phi$ is predominantly negative, and outgoing waves generally exhibit larger magnitudes than ingoing waves.
\label{fig:phaseshift}}
\end{figure}

The resulting phase shifts $\delta \phi$ relative to the quiet Sun are shown in Figure \ref{fig:phaseshift}. $\delta \phi$ is predominantly negative for both ingoing and outgoing waves, indicating that waves traveling from or into the sunspot have shorter travel times. Outgoing waves generally exhibit larger $\delta \phi$ than ingoing waves, suggesting the probable influence of subsurface flows or other systematic effects \citep{Zhao:2020aa}. For traveling angles $\theta$ = 150°, 180°, and 210°, measurements at distances shorter than the sunspot radius ($\approx17.5$~Mm) are unreliable due to the lack of sufficient quiet-area pixels inside the corresponding arcs. These data points are excluded from the analysis. In addition, outgoing‑wave measurements at distances below 13 Mm for all traveling angles are strongly affected by the low signal-to-noise ratio of the outgoing-wave branch and are similarly discarded. Short travel-distance ingoing waves are also affected, but the relatively high signal-to-noise ratio of the ingoing-wave branch makes the results less noisy, therefore they are not excluded but still need to be treated with caution.

\begin{figure}[htb!]
\plotone{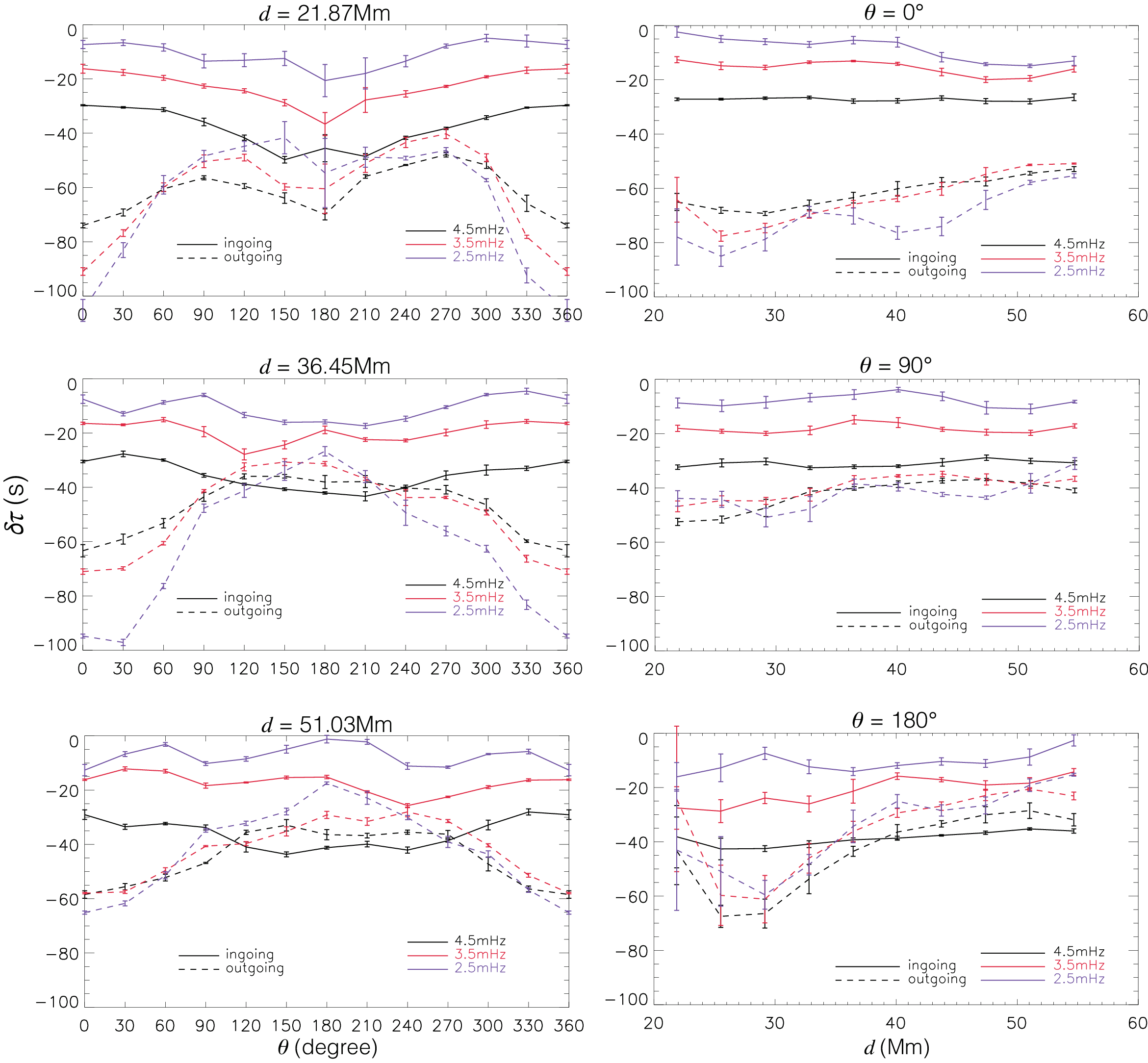}
\caption{Measured time shifts as functions of traveling angle $\theta$ and distance $d$. Solid and dashed lines indicate the results of ingoing and outgoing waves, respectively; purple, red and black lines indicate the results of 2–3~mHz, 3–4mHz and 4–5mHz (denoted as 2.5~mHz, 3.5mHz and 4.5mHz) respectively.
\label{fig:timeshift}}
\end{figure}

$\delta \phi$ decreases with frequency. 
Positive phase shifts are found at shorter travel distances and lower frequencies, more noticeably in the ingoing waves. Similar measurements were made in \citet{Braun:2008aa} These positive phase shifts probably relate to upward propagation due to mode-conversion \citep{Schunker&Cally2006MNRAS,Rajaguru:2010aa}. The detailed interpretation of such measurements is not yet well understood and is beyond the scope of this study. We will primarily focus on the negative phase shifts seen in Figure \ref{fig:phaseshift}, and positive phase shifts are found to have little influence on the travel-time shift result of the following rebin process. When frequency approaches about 5~mHz, in most panels the $\delta \phi$ reaches $-\pi$ which is the lower boundary of $\arg$ function (in Figure \ref{fig:phaseshift} the color scale is limited to $[-0.75\pi,0.75\pi]$ for the convenience of display), therefore phase wrapping occurs and ambiguities are introduced in the determination of travel times. We restrict our analysis to frequency range 2–5 mHz. $\delta \phi$ is averaged over three frequency bands: 2–3 mHz, 3–4 mHz, and 4–5 mHz, referred as 2.5 mHz bin, 3.5 mHz bin, and 4.5 mHz bin hereafter%, with representative frequencies of 2.5, 3.5, and 4.5 mHz, respectively
. Spatial rebinning of 6.5~Mm in travel distance is also applied. The resulting travel-time shifts $\delta\tau$ as functions of traveling angle, frequency, and distance are shown in Figure~\ref{fig:timeshift}. Error bars indicate the standard deviation within each distance bin.

Both ingoing and outgoing waves exhibit predominantly negative $\delta\tau$. The magnitude of the reduction reaches up to $\sim$40~s in average at short travel distances and gradually decreases with increasing distance. As shown in the left panels of Figure \ref{fig:timeshift}, both ingoing and outgoing $\delta\tau$ are approximately symmetric about $\theta$ = 180°, suggesting a near‑axisymmetric structure of the sunspot. For results of the ingoing waves, the magnitude shows only a weak dependence on traveling angle and distance. In contrast, outgoing $\delta\tau$ is larger in magnitude---exceeding one minute in some cases---and exhibit a stronger angular dependence. The magnitude generally decreases with increasing angle from 0° to 180°. Furthermore, we note that a local peak near 180° is observed at short distances, in agreement with previous findings \citep{10.1093/mnras/stv506}.

The angular behavior of the outgoing waves likely reflects the influence of the magnetic field. For traveling directions near $\theta = 180^\circ$, wave paths traverse a larger fraction of the sunspot near-surface region, where the impact of magnetic fields on wave speed is enhanced. As the traveling direction deviates from the radial direction, acoustic waves are more likely to undergo mode conversion into multiple types of MHD waves, probably resulting in phase shifts. The observed angular profiles thus represent a combination of these effects.

At larger traveling distances, the magnitude of outgoing $\delta\tau$ tends to decrease, particularly for directions near 0° and 180°. In contrast, the ingoing shifts show relatively little dependence on distance. 
Ingoing $\delta\tau$ increases in magnitude with frequency. $\delta\tau$ at 4.5 mHz is nearly three times larger than that at 2.5 mHz. For outgoing waves, however, the frequency dependence is less systematic.

In summary, three properties of the measured travel times are particularly notable. First, both ingoing and outgoing waves generally exhibit shorter travel times than those measured in the quiet Sun. Second, a systematic asymmetry exists between ingoing and outgoing waves, with outgoing travel times substantially shorter than the ingoing travel times. Third, the magnitude of $\delta\tau$ depends on the traveling angle, distance, and frequency. The results are similar to those previously obtained in other studies \citep[e.g.,][]{Liang2013,10.1093/mnras/stv506,Duvall2018}. These properties provide important clues to the subsurface structure and dynamics of the sunspot, and we attempt to interpret these measurements using numerical simulations and forward modeling in Section \ref{sec:simulation}.

\section{Numerical Models and forward modeling of travel times}\label{sec:simulation}
\subsection{MHS Model and numerical simulations}\label{subsec:simulat}

To further investigate the subsurface structure of the sunspot and to aid the interpretation of the measured travel times, we apply the same time–distance helioseismic analysis to Doppler velocity data generated from an MHS sunspot model. The model and simulation setup are the same as \citet{Felipe2017}, in which acoustic waves are excited by sources randomly distributed in horizontal directions and located at a depth of 0.15~Mm beneath the quiet-Sun photosphere layer (the sources inside the sunspot are moved deeper to form a constant temperature surface). The characteristics of wave excitation follows \citet{Parchevsky:2008aa}, and the resulting wave field is used to construct synthetic Dopplergrams.

The model incorporates various characteristic of sunspots, including magnetic field, Wilson depression, and sound-speed perturbations (a single near-surface layer of reduced sound speed $c_\mathrm{s}$). The maximum photospheric magnetic field strength in the model umbra is set to approximately 1500 G, and the radius of the modeled sunspot is approximately\footnote{we simply define the sunspot boundary as where the magnetic field is $1/10$ of the maximum to obtain the radius.} 17.5 Mm, both consistent with the observed sunspot. %Both the magnetic and thermal structures are included in the model. 
However, the MHS model does not contain plasma flows. The deepest part of the Wilson depression in the model is approximately 550 km.

According to \citet{Felipe_2016}, the vertical velocity at the height corresponding to the optical depth of $\tau=0.01$ is taken as the synthetic Doppler velocity signal. The same data analysis and time–distance measurement procedures described in Section~\ref{sec:Observations and time-distance measurement} are applied to the simulated Dopplergrams, allowing for a direct comparison between the simulations and observations.

\subsection{Eikonal Method of Forward Modeling}\label{subsec:The Eikonal Method}

In addition to the numerical simulations, we perform forward modeling of acoustic-wave travel times using ray tracing method based on eikonal approximation \citep{Weinberg:1962aa}. Travel times in sunspots are influenced by several factors, including plasma flows, sound-speed perturbations, magnetic fields, and Wilson depression. In addition to the interior physical properties-caused travel-time shifts, there is possibly a travel-time shift caused by that observations are made at different atmospheric heights, similar to the center-to-limb effect \citep{Zhao_2012,Chen_2018,Waidele:2023aa}. Disentangling these contributions through inversions alone is challenging; therefore, forward modeling provides a useful complementary approach to assessing whether a given sunspot model can qualitatively reproduce the observed travel-time behavior. 

Eikonal approximation assumes that the wavelength is much smaller than the characteristic length scales of the background medium, all wave variables depend on time as $e^{-i\omega t}$, and only phase variations of the wave are considered \citep{Khomenko_2009}. We perform calculations using two sunspot models. The first is the same MHS model described in Section~\ref{subsec:simulat}. To investigate the role of plasma flows, we also employ a model derived from MHD simulations presented in \citet{Braun:2012aa}, which uses the method of \citet{Rempel:2009sci,Rempel_2009apj}. The original 3D MHD model is temporally (27.3 hr) and azimuthally averaged around the sunspot central axis for an axisymmetrical 2D model. This model has a sunspot radius of approximately 13.5 Mm and includes realistic subsurface flow fields, the maximum photospheric magnetic field is approximately 1200~G. The wave sources and oscillatory field are not included in both models.

Wave propagation is computed using the dispersion relation for magnetoacoustic waves in a stratified, magnetized medium \citep{Moradi:2008aa,Khomenko_2009}:

\begin{equation}
\textit{D} \equiv \omega^4 - \omega^2 (c_\mathrm{s}^2 + v_\mathrm{A} ^2)({k_x}^2 + {k_z}^2) + c_\mathrm{s}^2 ({k_x}^2 + {k_z}^2)  %\times 
\left( v_{Ax} k_x + v_{Az} k_z \right)^2 - \omega_c^2 \left( \omega^2 - c_\mathrm{s}^2 {k_z}^2 \right) + c_\mathrm{s}^2 N^2 {k_x}^2 = 0
\label{eq:eq4}
\end{equation}

\noindent where $k_x$ and $k_z$ are the horizontal and vertical wavenumbers, respectively; $c_\mathrm{s}=\sqrt{\Gamma p/\rho}$ is the sound speed% from density, pressure and first adiabatic exponent $\Gamma$ 
, where $\Gamma$ is the first adiabatic exponent, $p$ is gas pressure and $\rho$ is density; $v_\mathrm{A} $ is the Alfvén speed, $N$ is the Brunt-Väisälä frequency, satisfying $N^2=g/H-g^2/c_\mathrm{s}^2$ and the isothermal cutoff frequency is $\omega_c=c_\mathrm{s}/2H$, where $H$ is the pressure scale height defined as $H = -p(\frac{\mathrm{d}p}{\mathrm{d}z})^{-1}$, and $g$ is gravitational acceleration. All background quantities (such as $c_\mathrm{s}$, $v_\mathrm{A} $, $H$, ..) are taken directly or calculated from the corresponding sunspot models. The ray paths in phase space are determined by integrating the Hamiltonian equations

\begin{equation} 
    \frac{\mathrm{d}x}{\mathrm{d}s} = \frac{\partial \textit{D}}{\partial k_x}
    \label{eq:eq5}
\end{equation}

\begin{equation} 
    \frac{\mathrm{d}z}{\mathrm{d}s} = \frac{\partial \textit{D}}{\partial k_z}
    \label{eq:eq6}
\end{equation}

\begin{equation} 
    \frac{\mathrm{d}k_x}{\mathrm{d}s} = -\frac{\partial \textit{D}}{\partial x}
    \label{eq:eq7}
\end{equation}

\begin{equation} 
    \frac{\mathrm{d}k_z}{\mathrm{d}s} = -\frac{\partial \textit{D}}{\partial z}
    \label{eq:eq8}
\end{equation}

The parameter $s$ is the ``distance'' along the wave ``path'' in the phase space% and has no direct significance\citep{Weinberg:1962aa}
. These equations are integrated numerically using a fourth-order Runge-Kutta method to trace fast magnetoacoustic mode. To calculate wave paths in the quiet Sun, the left and right side boundaries of the model are extended about 49~Mm each laterally to form an unperturbed quiet area.%reference region.

To isolate contributions from different physical effects, two sets of calculations are performed. In the first set of calculations, the magnetic and thermodynamic structures are retained while plasma flows are removed (for the MHD model). The wave travel time along a ray is computed using the phase integral $S(x,z)=\int (k_xdx+k_zdz)$ and the travel-time shift $\delta\tau$ is obtained by dividing the phase shift by the angular frequency $\omega$. In the second set of calculations, only the flow field is included. The travel-time shift due to flows is estimated using a first-order approximation \citep{Kosovichev:1996aa}, 

\begin{equation}
    \delta \tau=-\int \frac{1}{c_\mathrm{s}^2}(v_xdx+v_zdz)
\end{equation}

\noindent where $v_x$ and $v_z$ are the flow velocity components. The effect of flows on the ray path itself is neglected, and quiet-Sun ray paths are used. Under this approximation, ingoing and outgoing waves have equal-magnitude but opposite-sign time shifts, so the travel-time difference $\delta\tau_\mathrm{diff}$ between the ingoing and outgoing waves is 2$\delta \tau$. For the MHS model, only the non-flow calculations are performed, whereas for the MHD model both non-flow and flow-only calculations are carried out.

Forward modeling is performed for three representative frequencies: 2.5, 3.5, and 4.5 mHz, matching the frequency bands used in the observational analysis. For each travel distance at each angle, 10 ray paths are calculated, and their inner endpoints are located in $\sqrt{0.1}R, \sqrt{0.2}R, ...,R$ ($R$ is the sunspot radius, 17.5Mm for the MHS model and 13.5Mm for the MHD model) from the sunspot center, the final result is their average. For each case, rays are initialized at the lower turning point with $k_z=0$, and $k_x$ is determined from the dispersion relation (Equation \ref{eq:eq4}), the position of the initiation point is manually chosen to ensure that the resulting ray path has the needed horizontal travel distance. Travel time is integrated between the upper turning points of the ray path.

\subsection{Results and Interpretation}

We have obtained one set of travel-time measurements from the MHS simulation, and have obtained two sets of forward modeling results from MHS and MHD sunspot models, separately. In the measurements from MHS simulation data, the travel-time differences between ingoing and outgoing waves are found to be close to zero. This result is expected, as the MHS model contains no flow field, and neither magnetic fields nor thermal perturbations alone are theoretically predicted to produce systematic ingoing–outgoing asymmetries \citep[e.g.,][]{Moradi:2008aa}. This behavior is consistent with the non-flow forward-modeling results obtained from both the MHS and MHD models.

\begin{figure}[htb!]
\resizebox{1.0\textwidth}{!}{\plotone{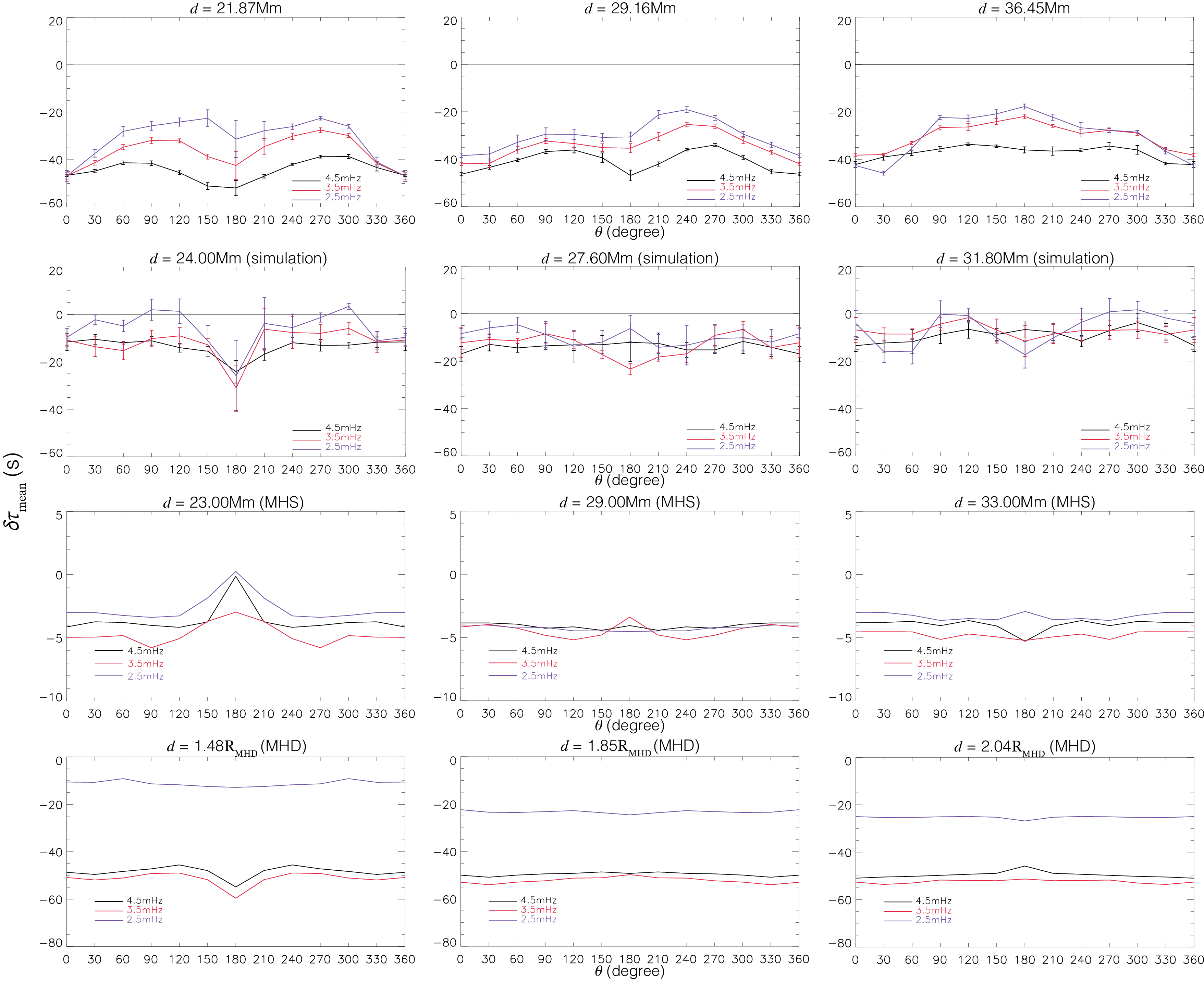}}
\caption{Angular dependence of measured and modeled mean travel-time shifts. Rows from top to bottom: results of observed sunspot, MHS model simulation, MHS ray-trace calculation, and MHD ray-trace calculation, respectively. The real sunspot and MHS model have radius of 17.5Mm, and the radius of MHD model is 13.5Mm, denoted as $R_\mathrm{MHD}$.
\label{fig:averageshift_angle}}
\centering
\end{figure}

\begin{figure}[htb!]
\centering
\resizebox{1.0\textwidth}{!}{\plotone{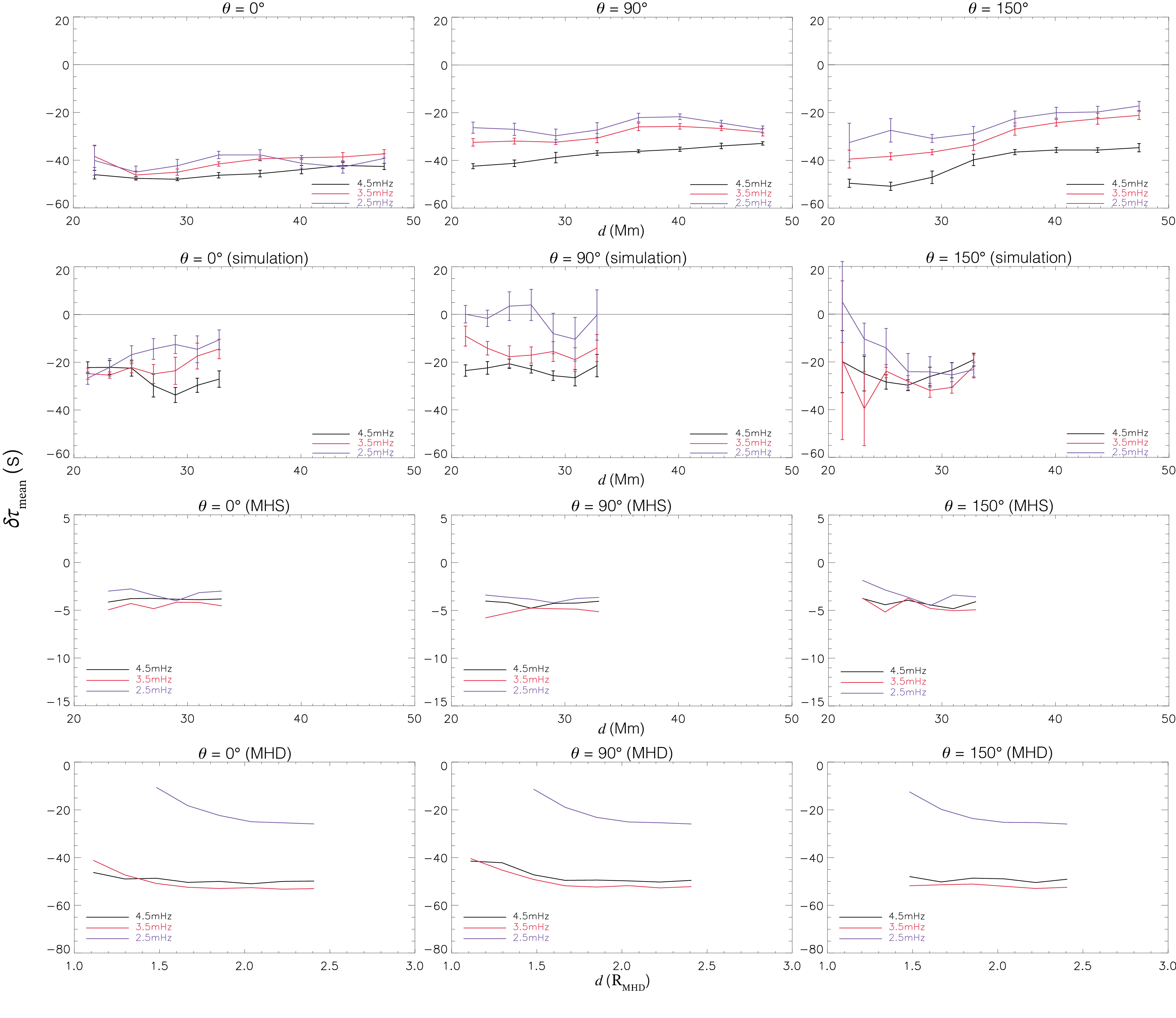}}
\caption{Distance dependence of measured and modeled travel-time shifts.
\label{fig:averageshift_dist}}
\end{figure}

We therefore focus on the mean travel-time shifts $\delta\tau_\mathrm{mean}$ of ingoing and outgoing waves, defined as the mean of the two, for both the observations and the no-flow simulation. These quantities primarily reflect the influence of magnetic fields, sound-speed perturbations, and Wilson depression. The resulting angle and distance dependencies are shown in Figures~\ref{fig:averageshift_angle} and~\ref{fig:averageshift_dist}.

For the observed sunspot, $\delta\tau_\mathrm{mean}$ is predominantly negative, with magnitudes of approximately 40 s, and exhibit approximate symmetry about $\theta=180$° (Figure~\ref{fig:averageshift_angle}). The simulation results are noisier due to the limited number of wave sources, but a similar axisymmetry is still evident. The overall magnitude of the simulated shifts is close to that reported by \citet{Felipe2017}, though smaller than the observed values. The MHS Eikonal modeling produces even smaller magnitudes, whereas the MHD model reproduces magnitudes closer to the observations, except at 2.5 mHz. \citet{Cally_2006} claimed that the WKB approximation used to derive the dispersion relation becomes worse as frequency decreases, which may explain the reduced magnitude at low frequency. It is hard for a sunspot model to precisely reproduce the structure of the observed sunspot, and we can only adjust the size and magnetic field of the MHS model, so it is reasonable that these results do not quantitatively match. However our simulation and modeling results can still reproduce some of the measurement profiles qualitatively, and thus providing insights about the interior structures of the studied sunspot. Both the MHS simulation measurements and MHD forward calculation match the observation to a certain degree, so it’s unlikely that the discrepancy of MHS modeling result is caused by only the MHS model or the ray-trace method.

The observed magnitude of $\delta\tau_\mathrm{mean}$ increases with frequency, which can be attributed to the Wilson depression and magnetic effects. In the surface layers where $v_\mathrm{A} / c_\mathrm{s}$  is relatively high, fast-mode speed increases \citep{Khomenko_2009,Felipe_2016}, causing a reduction in the wave travel time. Higher-frequency waves reflect from shallower layers with high $v_\mathrm{A} / c_\mathrm{s}$ %they are strongly accelerated
, so they have larger $\delta\tau$. Lower-frequency waves reflect from deeper layers before encountering the strong near-surface increase in $v_\mathrm{A} / c_\mathrm{s}$ , leading to smaller accelerations, therefore smaller magnitudes of $\delta\tau$. In contrast, the modeling results show larger shifts at 3.5 mHz than at 4.5 mHz, opposite to the observational trend, indicating limitations in the models.

The angular dependence of the $\delta\tau_\mathrm{mean}$ is broadly reproduced by the simulations and the forward modeling using the MHD model (Figure \ref{fig:averageshift_angle}). For relatively short travel distances (approximately 1.5 times the sunspot radius, see the left panels of Figure \ref{fig:averageshift_angle}), the magnitude has two local maxima, one occurring near $\theta$ = 0° and another near $\theta$ = 180°. The 180° local maximum arises because of two reasons. First, waves traveling in these directions re-enter the near-surface regions with high $v_\mathrm{A} /c_\mathrm{s}$, leading to enhanced fast-mode accelerations both at the beginning and the end, and the travel times are shortened. As the traveling angle deviates from these directions or as the travel distance increases, this effect gradually diminishes. The second reason is the $\mathbf{k}-\mathbf{B}$ angle. For 3.5 mHz and 4.5 mHz MHD forward modeling with $d<2R_\mathrm{MHD}$, we make a rough estimation of the angle between wave vector $\mathbf{k}$ and total magnetic field $\mathbf{B}$%\footnote{In practical, the total Alfvén speed is used instead of magnetic field, but we still refer to this angle as $\mathbf{k}-\mathbf{B}$ angle.}
. This relative angle is obtained by averaging the angle between $\mathbf{k}$ and $\mathbf{B}$ along the part of the ray path where $v_\mathrm{A} / c_\mathrm{s} > 0.1$, and restricted to 0°-90°. We found that the $\mathbf{k}-\mathbf{B}$ angle decreases from about 60° to 30° as $\theta$ varies from 0° to about 120°, but increases from about 30° to 60° rapidly as $\theta$ varies from 120° to about 180°. This can also explain the MHD forward modeling result because the phase speed of fast magnetoacoustic wave increases with $\mathbf{k}-\mathbf{B}$ angle from 0° to 90°.

The distance dependence of $\delta\tau_\mathrm{mean}$ is shown in Figure~\ref{fig:averageshift_dist}. In the observations, the magnitude decreases with distance, particularly for directions near $\theta$ = 180°. This trend reflects the reduced interaction of longer ray paths with near-surface magnetized layers. %This trend reflects the inclination of penumbral magnetic fields: when distance is large, wave paths intersect the magnetic field at steeper angles, becoming less aligned with the field line therefore the net acceleration is reduced. As angle approaches 180°, when distance increases, the disappearance of double-entrance of subsurface high $v_\mathrm{A} /c_\mathrm{s}$ layer also contribute to the large slope of decreasing. 
%One notable fact is that when travel distance is small (less than 28Mm for observed sunspot and MHS model), the decreasing trend in magnitude is less obvious and even reversed. For large $\theta$ cases, the similar behavior is reproduced in the simulation and MHS forward modeling results, although the usable range of travel distance is smaller than that of observed sunspot due to the size limit of simulation domain. %the amplitudes are smaller. Because of the size limit of simulation domain, the measurable range of travel distance is smaller than that of real sunspot, but a decreasing trend in amplitude can still be seen when distance reaches 30Mm. Although the MHS modeling results produce low amplitude, their distance dependence matches the measurements well, and also consistent with \citet{Khomenko_2009}. 
In contrast, the MHD model, which features a relatively small penumbra and more vertical fields, shows increasing magnitude with distance.%, as steeper ray paths align more closely with the magnetic field.%However, in the MHD modeling---where the sunspot has a relatively small penumbra and predominantly vertical fields---the amplitude of average time shifts increases with distance, probably because the wave path with steeper angle aligns better with the field line instead.

\begin{figure}[htb!]
\centering
\resizebox{1.0\textwidth}{!}{\plotone{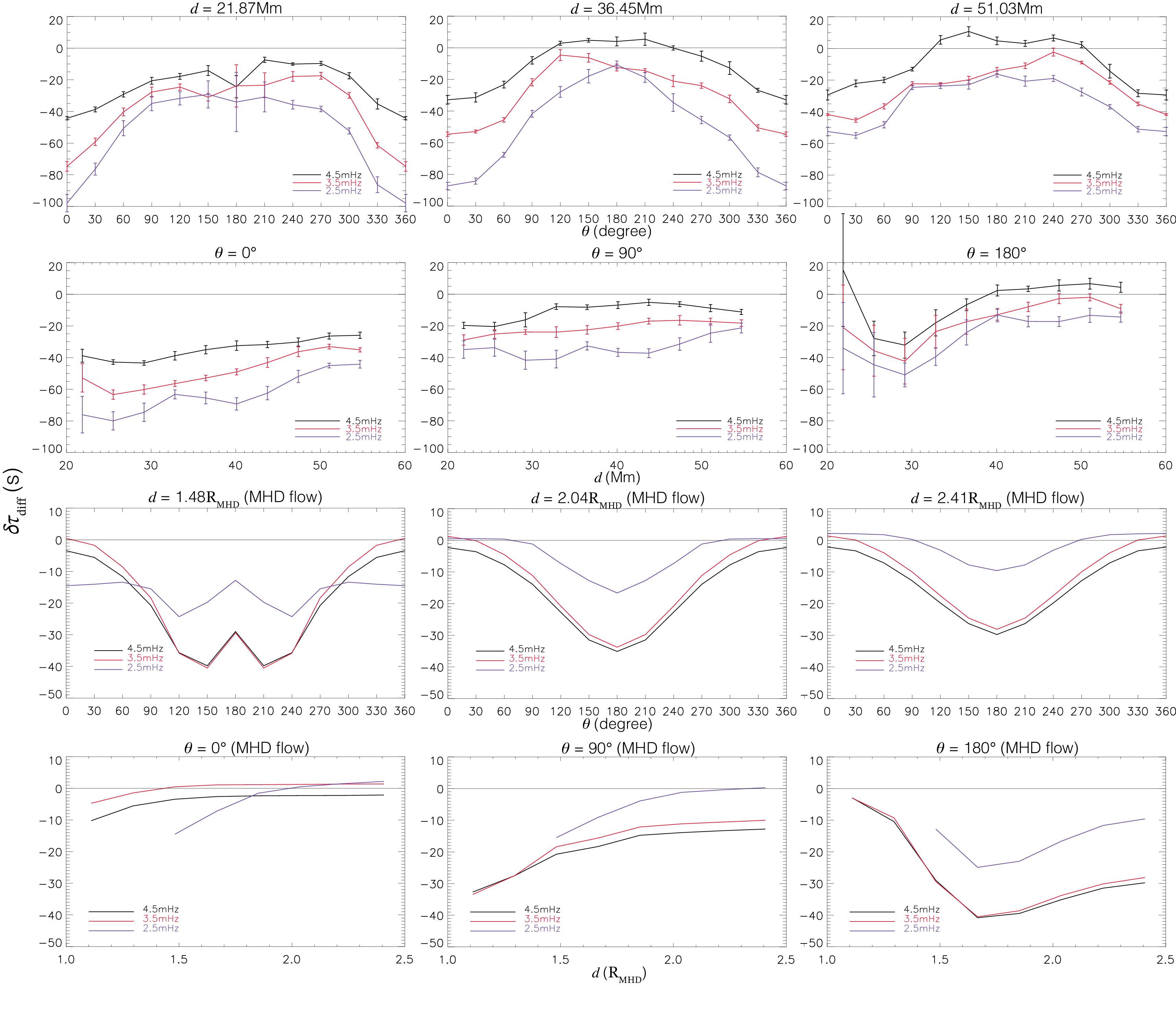}}
\caption{
Travel-time differences $\delta\tau_\mathrm{diff}$ between ingoing and outgoing waves as functions of angle and distance for three frequencies (2.5, 3.5, and 4.5 mHz). Upper two rows: observational measurements. Lower two rows: forward-modeling results including only plasma flows. 
\label{fig:timediff_flow}}

\end{figure}

To assess the contribution of subsurface flows to the observed outgoing-ingoing travel-time asymmetries, we compare the measured travel-time differences $\delta\tau_\mathrm{diff}$ (outgoing minus ingoing) with Eikonal modeling results that include only the flow field from the MHD model (Figure \ref{fig:timediff_flow}). 

\begin{figure}[htb!]
\centering
\resizebox{1.0\textwidth}{!}{\plotone{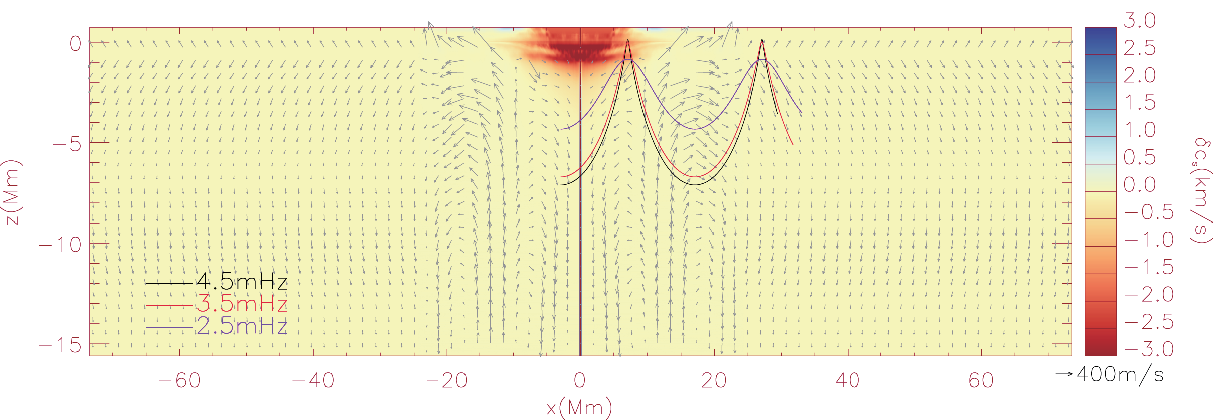}}
\caption{Diagram of the axisymmetric 2D MHD model, %passing through its center, 
showing the sound speed perturbation and flows. For the flow-only modeling cases, samples of wave paths for three frequencies are displayed (to the first-order approximation the wave paths for ingoing and outgoing waves have the same shape). The path calculation begins at the lower turning point and the travel time is integrated between two upper returning points.
\label{fig:model_and_raypath}}

\end{figure}

The observed $\delta\tau_\mathrm{diff}$ is generally negative, consistent with outward-directed flows beneath the sunspot, such as moat flow. The largest magnitudes occur near $\theta$ = 0°, where the flow is most closely aligned with the wave traveling direction. %As the angle increases, the component of the flow vector along the ray path decreases, resulting in a reduced amplitude. This trend continues until $\theta$ = 180°, where the path mostly opposes the flow direction and the time differences are minimized. 
When $\delta\tau_\mathrm{diff}$ caused by flows are combined with $\delta\tau_\mathrm{mean}$, they would cause $\delta\tau$ of ingoing waves being less sensitive to angle than that of outgoing waves. It is also notable that the magnitude of $\delta\tau_\mathrm{diff}$ decreases with frequency, contrary to $\delta\tau_\mathrm{mean}$. This is due to the waves of different frequencies have different ray paths for the same travel distance. Although waves of higher frequency have shallower upper turning points, the ray-calculation results (Figure \ref{fig:model_and_raypath}) show that the paths of lower frequency waves tend to be shallower overall. Other work \citep[e.g.,][]{Zhao:2020aa} also show that other factors such as Wilson depression would also cause frequency-dependent time difference, which will be discussed in Section~\ref{sec:Summary}. In our forward modeling, only flows are considered.

The overall magnitude of the modeled $\delta\tau_\mathrm{diff}$ is smaller than the observations, probably because the flow in the model is weaker than the observed sunspot. On the other hand, the angular dependence is reversed: in the model, the maximum occurs near $\theta$ = 180°. This discrepancy arises from the flow structure of the MHD model, which lacks an extended penumbral region with strong horizontal outflows and instead exhibits relatively strong subsurface inflows beneath the umbra \citep{Braun:2012aa}. As illustrated in Figure \ref{fig:model_and_raypath}, outgoing waves traveling near $\theta$ = 0° %direction, overlaid with sample ray paths for 2.5, 3.5, and 4.5 mHz waves. For each frequency, the ingoing and outgoing rays follow the same trajectory, due to the first-order calculation treatment.
initially traverse regions of upward flow that delay the waves, and the subsequent acceleration by horizontal outflows is insufficient to compensate for this effect. The net effect is negligible acceleration. By contrast, waves traveling near $\theta$ = 180° are accelerated by a more favorable alignment with the flows, resulting in a more apparent asymmetry between ingoing and outgoing waves.

Despite these discrepancies, the modeled distance dependence of $\delta\tau_\mathrm{diff}$ agrees qualitatively with the observations: the magnitude decreases with increasing distance, most rapidly near $\theta$ = 180°. The magnitude difference between $\theta$ = 0° and $\theta$ = 180° also decreases with distance. This behavior reflects the reduced sensitivity of deeper ray paths to localized near-surface flows.

Overall, these results demonstrate that while forward modeling can reproduce many qualitative features of the observed travel-time shifts and asymmetries, accurately capturing their angular and frequency dependence still remains a challenge. The forward modeling procedure may not capture all correct physics. Adjusting the MHS sunspot parameters, or adopting other sunspot models (e.g., those with more realistic magnetic and flow structures particularly in the penumbral region) may also lead to different results.

\section{Discussion and Summary}\label{sec:Summary}

Using time–distance helioseismology, we measure travel-time shifts of acoustic waves traveling into and out of a sunspot in AR 11312. The analysis was carried out over a wide range of travel distances and azimuthal angles relative to the sunspot’s radial direction, allowing us to characterize both mean travel-time shifts and asymmetric outgoing–ingoing travel times. 
The measurements reveal several systematic properties of the acoustic travel times. First, both ingoing and outgoing waves generally exhibit shorter travel times than those measured in the quiet Sun, especially along sunspot's radial direction ($\theta=0$°) and $\theta=180$° for short-distance cases. The magnitudes of measured travel-time shifts reach up to approximately 50~s%, indicating the impact of the Wilson depression, magnetic field and sound-speed perturbation
. Second, a clear asymmetry exists between ingoing and outgoing waves. Outgoing waves have shorter travel times, and the out-in time difference reaches a maximum of 60~s along the radial direction. Third, the magnitude of the travel-time shifts depends not only on traveling angle, but also on travel distance and frequency: the magnitudes of outgoing time shifts decrease with travel distance while ingoing time shifts show little distance dependence; the magnitudes of ingoing time shifts significantly increase with frequency but outgoing time shifts do not have a clear frequency dependence. 

To interpret these measurements, we performed numerical simulations of wave propagation in a magnetohydrostatic (MHS) sunspot model following \citet{Felipe_2016}, and applied the same time–distance analysis on the resulting synthetic Dopplergrams. In parallel, we carried out forward modeling using ray tracing under the eikonal approximation, based on both the MHS model and a magnetohydrodynamic (MHD) sunspot model that includes realistic flow fields \citep{Rempel_2009apj,Rempel:2009sci,Braun:2012aa}. Below we summarize the physical interpretation of measurement results based on comparisons with numerical simulations and forward modeling.

First, what has caused the negative travel-time shifts of both ingoing and outgoing waves? Comparisons with numerical simulations and forward modeling indicate that this reduction can be largely explained by a combination of magnetic fields, sound-speed perturbations, and the Wilson depression, because the results based on the MHS model which contains these factors qualitatively match the measurements. The Wilson depression reduces the geometric path length of acoustic waves traveling between the sunspot and the surrounding region, leading to shorter travel times. According to \citet{Khomenko_2009} and \citet{Felipe_2016}, magnetic fields modify the fast mode wave speeds in the near-surface layers of the sunspot. Although the simulations reproduce the qualitative behavior of the observations, the magnitude of the travel-time shifts is somewhat smaller than observed. Several factors may contribute to this magnitude discrepancy. Firstly, one may consider that the MHS model includes only a single near-surface layer of reduced wave speed, whereas helioseismic inversions commonly infer a two-layer structure beneath sunspots, with a deeper region of enhanced wave speed  \citep{Kosovichev2010LocalHO,Moradi:2010aa} which may significantly affect travel times. However, \citet{Felipe_2016} showed that when only thermal structures are included, a sunspot model with one single layer of reduced sound speed can still produce significant negative travel-time shifts, so inversions of these measurements may not be accurate, especially when assuming that the travel-time shifts are due to changes in the wave speed. This indicates that the double-layer assumption may not explain the travel-time shift mismatch in our work, and the contribution of other thermal-related structures such as the Wilson depression may overweigh that of wave-speed perturbation. %Given that the waves analyzed here travel horizontally over distances exceeding 20 Mm, they are expected to sample deeper layers where this positive perturbation may significantly affect travel times. 
Secondly, uncertainties in the Wilson depression may also play a role. \citet{10.1093/mnras/stv506} showed that a change in the Wilson depression of 100 km can modify travel times by several seconds, depending on frequency. The Wilson depression in the MHS model is fixed at 550 km at the center of the sunspot, while the true depression of the observed sunspot is not independently constrained; if there are intensity measurements of Fe \textsc{i} line, one may try to invert the thermal structure and to estimate Wilson depression in a future work. In summary, the combined deviations in sound-speed structure, magnetic configuration, and mainly Wilson depression likely account for the remaining mismatch between the modeled and observed mean travel-time shifts.

Second, the measurements reveal a systematic asymmetry between ingoing and outgoing travel times. In the numerical simulations and forward modeling, magnetic field and thermal perturbations alone produce symmetric travel-time shifts with negligible ingoing–outgoing differences, and the observed asymmetry is partially explained by subsurface flows. When only plasma flows are included in the MHD forward modeling, the resulting travel-time asymmetries reach half of the observations, confirming that subsurface flows can efficiently contribute to the travel time asymmetry. The difference between the modeling and observation results could be explained by the relatively weak flow of the model, but other factors may also contribute to this difference, which would be discussed later. However, the modeled angular dependence is reversed: in the observations, the largest asymmetries occur near $\theta$ = 0°, whereas in the model they peak near $\theta$ = 180°. This inconsistency arises from limitations of the adopted MHD model, which features a relatively small penumbra and weak horizontal outflows, as well as strong subsurface inflows beneath the umbra. These characteristics slow down outgoing waves traveling radially outward and lead to an incorrect angular signature. The comparison between observations and modeling therefore indicates that realistic, extended penumbral flow structures are essential for reproducing the measured angular dependence of travel-time asymmetries. % Models lacking sufficiently strong and spatially extended horizontal outflows cannot adequately capture the observed behavior.
%In the modeling, when only the plasma flow is considered in the MHD model, the average magnitude of the travel-time asymmetry agrees with the measurements, but the angular dependence is reversed. In real measurements, time differences peak near $\theta$ = 0°, while in the model they peak at $\theta$ = 180°. This discrepancy is attributed to the limited spatial extent and weakness of horizontal outflows in the model’s penumbra. The model features relatively large upward inflows in the umbral subsurface, which hinder outgoing waves at small propagation angles. These results show that measured sunspot includes strong, extended outflows, which are necessary for a model to reproduce measured frequency dependence of time differences. 
In addition to plasma flows, other mechanisms may also contribute to travel-time asymmetries. \citet{Zhao:2020aa} reported a systematic effect in time-distance helioseismology, finding that for the ray path whose two endpoints were observed at different atmospheric heights (one endpoint in a higher atmospheric layer using SDO/HMI line-core intensity, and the other in a lower atmospheric layer using SDO/HMI continuum intensity), a travel-time difference up to 2 minutes is measured between high-to-low case and low-to-high case. They also stated that this effect is similar in nature to the center-to-limb effect \citep{Zhao:2013aa,Chen_2018}, which is measured in Dopplergrams in quiet-Sun regions from disk center to limb where the effective atmospheric-height changes modestly and has travel-time difference of a few seconds. Relating to our current work, the effect reported by \citet{Zhao:2020aa} has the setup analogous to the sunspot case in the sense that the in-sunspot pixel may sample a lower atmospheric height, while the outer quiet-Sun pixel samples a higher atmospheric height. This suggests that this effect could be relevant to the observed $\delta\tau_\mathrm{diff}$, although a quantitative estimate would require modeling efforts and knowledge of the atmospheric-height differences. Following \citet{Zhao:2020aa}, we assume that this effect introduces relatively small time differences that depend primarily on frequency and travel distance, but not strongly on azimuthal angle, especially for large distances. A detailed investigation of this effect is beyond the scope of the present study.

Third, the travel-time shifts also exhibit strong dependences on traveling angle and frequency. Studying the origin of these dependences is also worthwhile. These dependences likely arise from the anisotropic interaction between acoustic waves and the magnetic field of the sunspot. Waves traveling approximately along the radial direction of the sunspot interact more strongly with the magnetic field, producing travel-time shifts with larger magnitudes. In our numerical simulations and forward modeling, the mean time shifts are less sensitive to the traveling direction. This discrepancy warrants further investigation. According to the forward modeling, the frequency dependence of the travel-time shifts may be related to the different heights and depths sampled by waves of different frequencies. Higher-frequency waves are more sensitive to the near-surface layers where magnetic effects are strongest. The reason why ingoing waves exhibit larger frequency dependence remains uncertain.

Table~\ref{tab:factors} summarizes the expected impacts of magnetic fields, sound-speed perturbations, plasma flows, and the Wilson depression on helioseismic travel times, based on the combined observational and modeling results. The results presented here demonstrate that directional time–distance measurements provide valuable diagnostics of sunspot structure and dynamics. By combining observations with numerical modeling, it is possible to separate the contributions from magnetic fields, thermal perturbations, and flows. Future studies using more realistic sunspot simulations % and longer observational time series
may further improve our understanding of wave propagation in strongly magnetized regions of the Sun.

%Requires: \usepackage{graphicx}
\begin{table}[h]
    \caption{Summary of the expected impacts of various factors on helioseismic travel-time measurements, based on observational and modeling results. All factors are expected to cause time shifts to change with frequency because waves of different frequencies have different wave paths.}
    \begin{tabular}{|l|c|c|c|}
        \hline
        \textbf{Factor} & \textbf{Travel time} & \textbf{Time shifts change} & \textbf{Time shifts change}  \\
        & \textbf{differences} & \textbf{with angle} & \textbf{with frequency}  \\

        \hline
        Magnetic field & No & Yes & Yes  \\
        \hline
        Sound-speed perturbation & No & Yes & Yes  \\
        \hline
        Flow field  & Yes & Yes & Yes  \\
        \hline
        Wilson depression  & Yes & No & Yes  \\
        \hline
    \end{tabular}
    
    \label{tab:factors}
\end{table}

\section*{Acknowledgement}
H.L. and H.T are supported by the National Natural Science Foundation of China (Grant No.12425301), the National Key R\&D Program of China (Grant No.2022YFF0503800), the Specialized Research Fund for State Key Laboratory of Solar Activity and Space Weather and China’s Space Origins Exploration Program.

TF and EK acknowledge grants PID2021-127487NB-I00 and PID2024-156538NB-I00, funded by MCIN/AEI/ 10.13039/501100011033 and by “ERDF A way of making Europe”. TF also acknowledges grants CNS2023-145233 and RYC2020-030307-I, funded by MICIU/AEI/10.13039/501100011033.

We also thank Taoni Bao of Peking University for his help in improving English of this paper.

\bibliography{sample631}{}
\bibliographystyle{aasjournal}



\end{document}